\documentclass[oldversion]{aa}  
\usepackage{natbib}
\usepackage{graphicx}
\usepackage{rotating}
\usepackage{txfonts}
\newcommand{\ms}{$\,$M$_\mathrm{\odot}$}

\newcommand{\be}{\begin{equation}}
\newcommand{\ee}{\end{equation}}
\newcommand{\stars}{{\sc stars}}

\newcommand{\gr}{\bigtriangledown_\mathrm{rad}}
\newcommand{\gad}{\bigtriangledown_\mathrm{ad}}

\newcommand{\deltaov}{\ensuremath{\delta_\mathrm{ov}}}

\begin{document}

   \title{Confronting uncertainties in stellar physics: \\ calibrating convective overshooting with eclipsing binaries}
	\titlerunning{CUSP -- calibrating convective overshooting}

   \author{R. J. Stancliffe\inst{1} \and L. Fossati\inst{1} \and J.-C. Passy\inst{1} \and F. R. N. Schneider\inst{1,2}}

   \institute{Argelander-Institut f\"ur Astronomie, University of Bonn, Auf dem H\"ugel 71, D-53121 Bonn, Germany
   \and Department of Physics, University of Oxford, Denys Wilkinson Building, Keble Road, Oxford OX1 3RH, U.K.}

   \date{}

\abstract{As part of a larger program aimed at better quantifying the uncertainties in stellar computations, we attempt to calibrate the extent of convective overshooting in low to intermediate mass stars by means of eclipsing binary systems. We model 12 such systems, with component masses between 1.3 and 6.2\ms,  using the detailed binary stellar evolution code \stars, producing grids of models in both metallicity and overshooting parameter. From these, we determine the best fit parameters for each of our systems. For three systems, none of our models produce a satisfactory fit. For the remaining systems, no single value for the convective overshooting parameter fits all the systems, but most of our systems can be well described with an overshooting parameter between 0.09 and 0.15, corresponding to an extension of the mixed region above the core of about 0.1-0.3 pressure scale heights. Of the nine systems where we are able to obtain a good fit, seven can be reasonably well fit with a single parameter of 0.15. We find no evidence for a trend of the extent of overshooting with either mass or metallicity, though the data set is of limited size. We repeat our calculations with a second evolution code, {\sc mesa}, and we find general agreement between the two codes. For the extension of the mixed region above the convective core required by the {\sc mesa} models is about 0.15-0.4 pressure scale heights. For the system EI Cep, we find that {\sc mesa} gives an overshooting region that is larger than the \stars\ one by about 0.1 pressure scale heights for the primary, while for the secondary the difference is only 0.05 pressure scale heights.}

   \maketitle
%

\section{Introduction}

Using observations to constrain stellar evolution models is one of the primary drivers of stellar astrophysics research. Nevertheless, it is not  straightforward to directly link observations and theory in a fully consistent manner, properly taking into account the uncertainties, particularly the theoretical ones. As a result, observers often consider their preferred set of stellar evolution tracks as intrinsically correct, and only recently comparisons between the stellar parameters gathered from different sets of models have begun to be performed, mostly for low-mass stars \citep[e.g.][]{2011A&A...530A.138C}. This ``exercise'' should be regularly performed, but in order to do that, particularly for a large number of stars, one needs to develop a dedicated automatic tool.

To this end, we have set up the ``Confronting Uncertainties in Stellar Physics' (CUSP) project, the aim of which is to better quantify the impact of theoretical uncertainties in the use of stellar evolution models to determine fundamental stellar parameters from observables. We intend to make use of {\sc bonnsai}\footnote{The {\sc bonnsai} web-service is available at {\tt www.astro.uni-bonn.de/stars/bonnsai}.} \citep{2014arXiv1408.3409S}, a publicly available tool, which allows one to derive stellar parameters (e.g., mass, radius, age) from a set of observational parameters (e.g., effective temperature, surface gravity, rotational velocity), properly accounting for the observational uncertainties. We will then explore the impact of both observational and theoretical uncertainties in the analysis of particular astrophysical problems which are directly linked to the estimation of stellar fundamental parameters on the basis of evolutionary tracks, such as the derivation of the masses and radii of transiting exoplanets.

In its current status, {\sc bonnsai} contains sets of evolutionary tracks only for intermediate to massive stars (M$>$5\,M$_\odot$) calculated for three different metallicities: Galactic, LMC and SMC \citep[][]{2011A&A...530A.115B,2015A&A...573A..71K}. We aim at extending the model database to lower mass stars from 0.8 to 10\ms, and for a variety of different metallicities. This extension of {\sc bonnsai} will be made by calculating new large grids of models initially with two different codes \stars\ \citep{1971MNRAS.151..351E,2009MNRAS.396.1699S} and {\sc mesa} \citep{2011ApJS..192....3P}, with the intention to add further grids calculated with other codes. However, the first step in the creation of these grids is the calibration of certain parameters. In this work, we focus on the issue of convective overshooting.

One of the key uncertainties in the evolution of main sequence stars is the size of the convective core. It is widely accepted that models based on either the Schwarzchild or Ledoux criterion produce cores that are too small to match observations and hence some form of overshooting must be applied to stellar models \citep[e.g.][]{1991A&AS...89..451M}. By overshooting we simply mean that the chemically mixed region in the star's core has been extended beyond the convective boundary predicted from standard stellar theory. This additional mixing could be caused by any number of phenomena, and is not necessarily related to the motion of material driven by convection `overshooting' the formal convective boundary.

Convective overshooting is a key component of canonical stellar models but it requires calibration. Many possible methods for calibration exist,  including the fitting of isochrones to stellar cluster colour-magnitude diagrams \citep[e.g.][]{2006ApJS..162..375V}. More recently, asteroseismology has opened up a new avenue for calibration of mixing properties in stellar interiors \citep[e.g.][]{2013ApJ...766..118M,2014ApJ...787..164G}. \citet{aerts2015} presents 16 OB dwarfs for which asteroseismic determinations of the extent of overshooting have been made. Similar analyses have also been carried out by \citet{2012A&A...539A..90N} and \citet{2014MNRAS.442..616T}. 

Here we focus only on calibrations using binary systems. \citet{1997MNRAS.285..696S} used $\zeta$ Aurigae-type systems to attempt to calibrate the extent of overshooting. They could find adequate fits to their systems (with masses between 2.5 and 6.5\ms) using overshooting equivalent to 0.24 - 0.32 pressure scale heights. Subsequently, \citet{1997MNRAS.289..869P} used the same evolutionary code and overshooting prescription to look at 49 eclipsing binary systems taken from the compilation of \citet{1991A&ARv...3...91A}. They found that models with and without overshooting could adequately fit the observations of the majority of the systems. However, for three systems (namely AI Hya, WX Cep and TZ For) models with enhanced mixing provided a better fit.

Further attempts to calibrate convective overshooting have been made by \citet{2007A&A...475.1019C}. He used 13 double-lined eclipsing binaries covering a range of evolutionary states and masses (from 1.3 to nearly 30\ms). Moderate amounts of overshooting of around 0.2 times the pressure scale height were found to best fit the data, with little evidence for a mass dependency. This is in contrast to the earlier work of \citet{2000MNRAS.318L..55R}, who suggested a mass dependence for overshooting may exist, based on a sample of 8 stars between 2 and 12\ms. More recent work by \citet{2014ApJ...787..127M}, using four eclipsing binaries in the mass range 1.3-3.6\ms, also finds no evidence for a mass dependency to overshooting, though the mass range is much smaller than the two studies mentioned above.

In this work, we revisit the issue of calibrating convective overshooting using eclipsing binaries. Ultimately, the aim is to arrive at a reliable determination that can be used for the computation of large grids of low-mass stellar models for use with the {\sc bonnsai} tool.

\section{Stellar models}

Computations in this work were made using the \stars\ stellar evolution code originally developed by \citet{1971MNRAS.151..351E} and updated by many authors \citep[e.g.][]{1995MNRAS.274..964P}. The code is freely available for download from \texttt{http://www.ast.cam.ac.uk/$\sim$stars}. This code solves the equations of stellar structure and chemical evolution in a fully simultaneous manner, iterating on all variables at the same time in order to converge a model \citep[see][for a detailed discussion]{2006MNRAS.370.1817S}. The version employed here is that of \citet{2009MNRAS.396.1699S} which was developed for doing binary stellar evolution. The code treats all forms of mixing by means of a diffusive formalism \citep{1972MNRAS.156..361E}.

\citet*{1997MNRAS.285..696S} describe the implementation of overshooting in the code. Rather than applying an extension to the convective region that is some fraction of a pressure scale height, this implementation makes an adjustment directly to the convective criterion (in this case the Schwarzchild criterion). A region is determined to be convectively unstable if $\gr > \gad - \delta$, where 
\be
\delta = {\delta_\mathrm{ov}\over 2.5 + 20\zeta +16\zeta^2}
\ee
with $\zeta$ being the ratio of radiation to gas pressure and \deltaov\ is a constant that must be determined (i.e. the overshooting parameter). This formalism ensures a smooth transition between systems with and without convective cores.

Each model is evolved from the pre main sequence using 999 mesh points\footnote{Select model sequences have been constructed using 1999 mesh points, and separately, twice the number of timesteps. The evolutionary tracks are indistinguishable from the ones presented. We are therefore satisfied that the computations are numerically converged.}. The mixing length $\alpha$ is set to 2.0, based on calibration to a Solar model. No mass loss is included. In addition, each binary is placed in a wide orbit so that there is no interaction between the components -- we do not attempt to reproduce the observed orbital period. Similarly, these models are non-rotating and we do not try to reproduce the observed rotational velocity for those systems where it has been measured. Rotational mixing would also act to increase the size of the chemically mixed region. Provided our target systems do not rotate rapidly, the use of non-rotating models should suffice. Each system is evolved to the point where the primary has comfortably exceeded the observed primary radius.

\section{Results}

\begin{figure}
\includegraphics[width=\columnwidth]{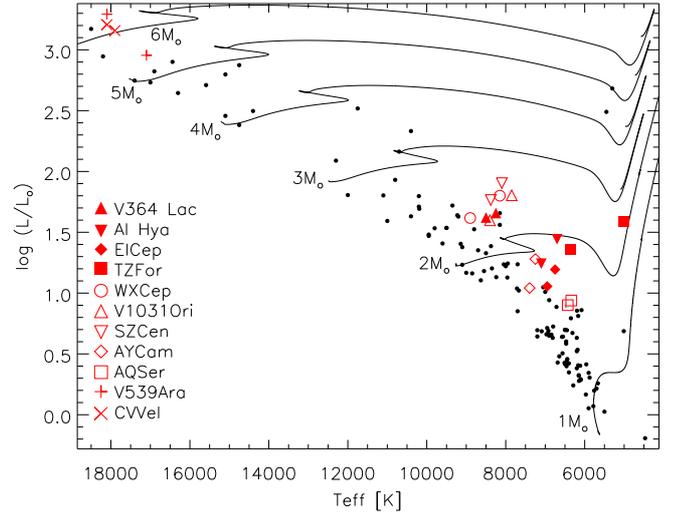}
\caption{Hertzsprung-Russell diagram showing eclipsing binaries from the \citet{2010A&ARv..18...67T} sample, with the systems we have selected shown in red. Stellar evolutionary tracks computed without overshooting are shown to give an approximate indication of mass and evolutionary state.}
\label{fig:selected_hr}
\end{figure}

We attempted to model 11 binary systems: V364 Lac, AI Hya, EI Cep, TZ For, WX Cep, V1031 Ori, SZ Cen, AY Cam, AQ Ser, V539 Ara and CV Vel. The properties of each of these systems are listed in Table~\ref{tab:obs}. These particular systems were selected from the sample of \citet*{2010A&ARv..18...67T}, which lists the known eclipsing binaries whose parameters have been determined to better than 3\%. We chose those systems which appear the most evolved in the Hertzsprung-Russell diagram (see Fig.~\ref{fig:selected_hr}), with both components of the system being clearly separated from the zero age main sequence. Evolved systems should be more sensitive to the effects of overshooting as they have been influenced by the process for longer. Where possible, we have avoided short period systems as tidal forces could have altered the stellar structure and evolution.

\begin{table*}
\begin{center}
\begin{tabular}{lcccccccccccc}
System & Period & Spectral & Mass & $\sigma$ & Radius & $\sigma$ & T$_\mathrm{eff}$ & $\sigma$  & $\log g$ & $\sigma$  & $\log L/\mathrm{L_\odot}$ & $\sigma$ \\
& (d) & type & (\ms) & & (R$_\odot$) & & (K) & & & & & \\
\hline
V539 Ara &  3.17 & B3V &   6.240 &  0.066 &  4.516 &  0.084 & 18100 & 500 & 3.924 & 0.016 &  3.293 &  0.051  \\
& & B4V  &      5.314 &  0.060 &  3.428 &  0.083 & 17100 & 500 & 4.093 & 0.021 &  2.955 &  0.055  \\
CVVel &  6.89 & B2.5V &   6.086 &  0.044 &  4.089 &  0.036 & 18100 &  500 & 3.999 & 0.008 &  3.207 &  0.049   \\
& & B2.5V &    5.982 &  0.035 &  3.950 &  0.036 & 17900 & 500 & 4.022 & 0.008 &  3.158 &  0.049  \\
 V364 Lac & 7.35 & A4m & 2.333 &  0.014 &  3.309 &  0.021 & 8250 & 150 & 3.766 & 0.005 & 1.658 & 0.032  \\
	         &  	   &  A3m & 2.295 & 0.024 & 2.986 & 0.020 & 8500 & 150 & 3.849 & 0.006 & 1.621 & 0.031  \\
AI Hya & 8.29 & F2m & 2.140 & 0.038 & 3.916 & 0.031 & 6700 & 60 & 3.583 & 0.006 & 1.443 & 0.017 \\
              &         & F0V & 1.973 & 0.036 &  2.767 & 0.019 & 7100 & 65 & 3.849 & 0.005 & 1.242 & 0.017 \\
EI Cep & 8.44 & F3V & 1.772 & 0.007 & 2.897 & 0.048 & 6750 & 100 & 3.763 & 0.014 & 1.194 & 0.030  \\	
	   &	       & F1m & 1.680 & 0.006 & 2.330 & 0.044 & 6950 & 100 & 3.929 & 0.016 & 1.056 & 0.030  \\
TZ For & 75.67 & G8III & 2.045 & 0.055 & 8.320 & 0.120 & 5000 & 100 & 2.908 & 0.013 & 1.589 & 0.037  \\
            &             & F7IV & 1.945 & 0.027 & 3.965 & 0.088 & 6350 & 100 & 3.531 & 0.018 & 1.361 & 0.033  \\
WX Cep &	 3.38 & A5V & 2.533 & 0.050 &  3.996 &  0.030 &  8150 &  250 & 3.638 & 0.005 &  1.801 &  0.054  \\
 & & A2V & 2.324 & 0.045 & 2.712 & 0.023 & 8900 & 250 & 3.938 & 0.006 & 1.617 & 0.050  \\
V1031 Ori &  3.41 & A6V & 2.468 &  0.018 &  4.323 &  0.034 &  7850 &  500 & 3.559 & 0.007 & 1.804 &  0.112 \\
 & & A3V & 2.281 & 0.016 & 2.978 & 0.064 & 8400 & 500 & 3.848 & 0.019 & 1.598 &  0.105  \\
SZ Cen & 4.11 & A7V &  2.311 &  0.026 &  4.556 &  0.032 &  8100 &  300 & 3.485 & 0.006 &  1.904 &  0.065  \\
& & A7V & 2.272 & 0.021 &  3.626 & 0.026 & 8380 &  300 & 3.676 & 0.006 & 1.765 & 0.062  \\
AY Cam &	 2.73 & A0V & 1.905 & 0.040 &  2.772 &  0.020 &  7250 & 100 & 3.832 & 0.004 & 1.280 &  0.025  \\
& & F0V & 1.707 &  0.036 &  2.026 &  0.017 &  7395 &  100 & 4.058 & 0.005 &  1.042 &  0.025  \\
AQ Ser & 1.69 & F5 & 1.346 &  0.024 &  2.281 &  0.014 &  6430 & 100 & 3.850 & 0.009 &  0.901 &  0.027  \\
& & F6 & 1.417 &  0.022 &  2.451 &  0.027 & 6340 & 100 & 3.810 & 0.012 & 0.939 & 0.042  \\
\hline
\end{tabular}
\caption{Parameters of the eclipsing binary systems used in this study. All the data comes from the compilation of \citet{2010A&ARv..18...67T}.}
\end{center}
\label{tab:obs}
\end{table*}

For each system, we run a grid of models across 4 metallicities (namely Z=0.01, 0.02, 0.03, 0.04) and with overshooting parameters $\delta_\mathrm{ov}$ from 0 to 0.30 in steps of 0.03. The helium content is presumed to vary as Y=0.25 + 15Z. To assess the quality of fit, for each timestep in the model sequence we calculate a goodness-of-fit via the formula:
\begin{equation}
P = \prod_i \exp \left( - {[x_i - \mu_i]^2 \over 2 \sigma_i^2}\right)
\end{equation}
where $x_i$ is the relevant parameter from the stellar model ($R$, or $T_\mathrm{eff}$) for each star in the system, and $\mu_i$ and $\sigma_i$ are the observed quantity and its error bar respectively. Note this means we are fitting both the primary and secondary simultaneously at each timestep. If our model fits perfectly we obtain P=1. We specifically use only $R$ and $T_\mathrm{eff}$ in this fit (and not surface gravity or luminosity) as these are the {\it directly measured} quantities. For the purposes of this study, we have not investigated the effects of uncertainties in the mass determinations.
\subsection{EI Cep}

A best fit for EI Cep is obtained with our Z$=0.02$ model with $\delta_\mathrm{ov}=0.15$ model, where we obtain P = 0.9066. In Fig.~2, we show the evolutionary tracks for both components of this system in the HR diagram, together with the observed characteristics of the system. Our best fit timestep is displayed by the crosses in the figure.

\begin{figure*}
\begin{center}
\includegraphics[width=\columnwidth]{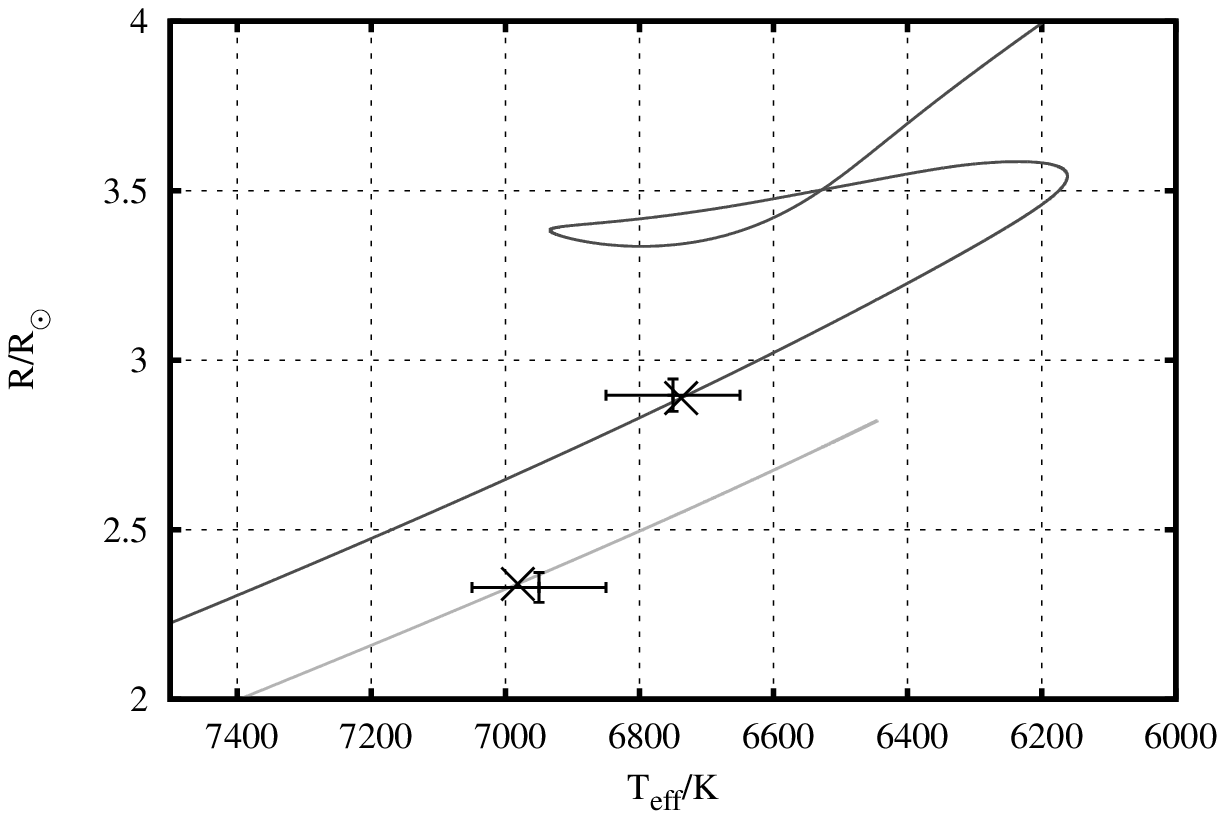}
\includegraphics[width=\columnwidth]{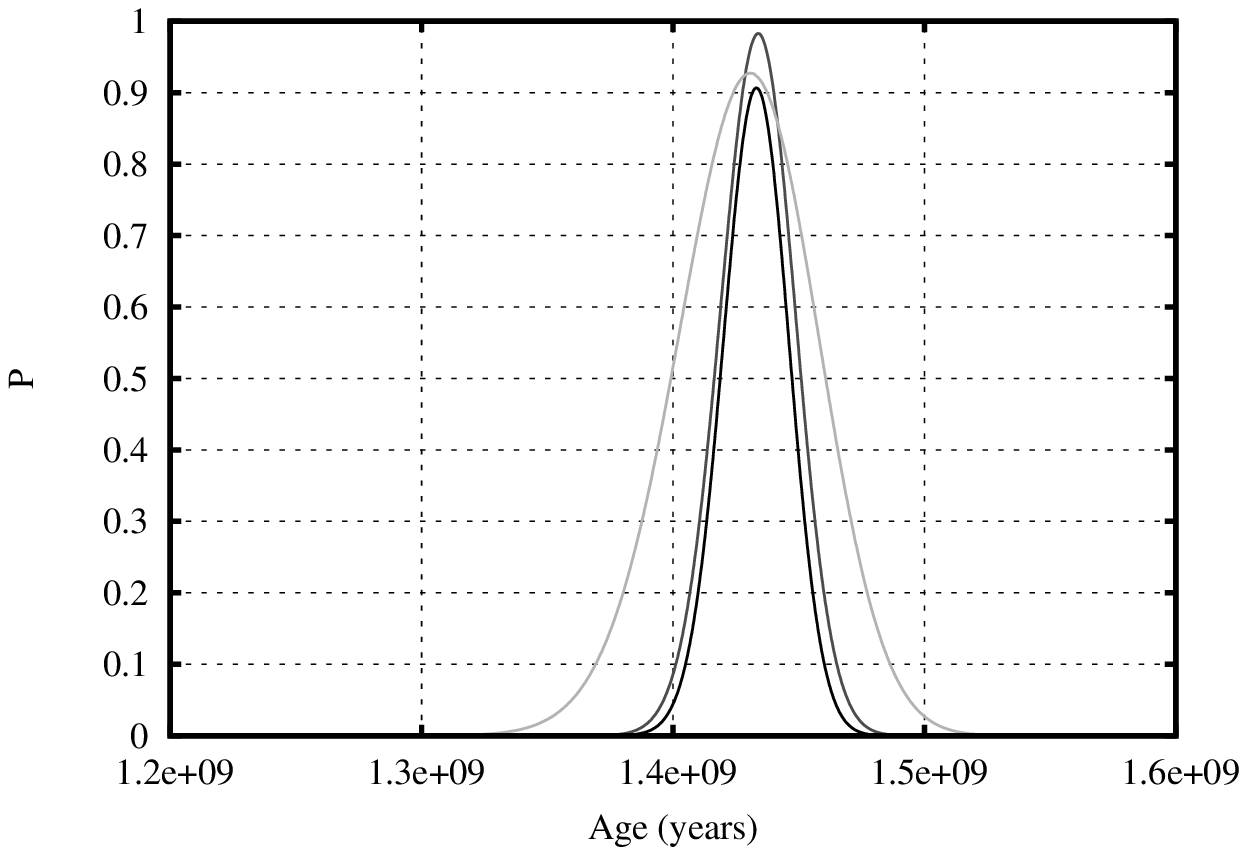}
\includegraphics[width=\columnwidth]{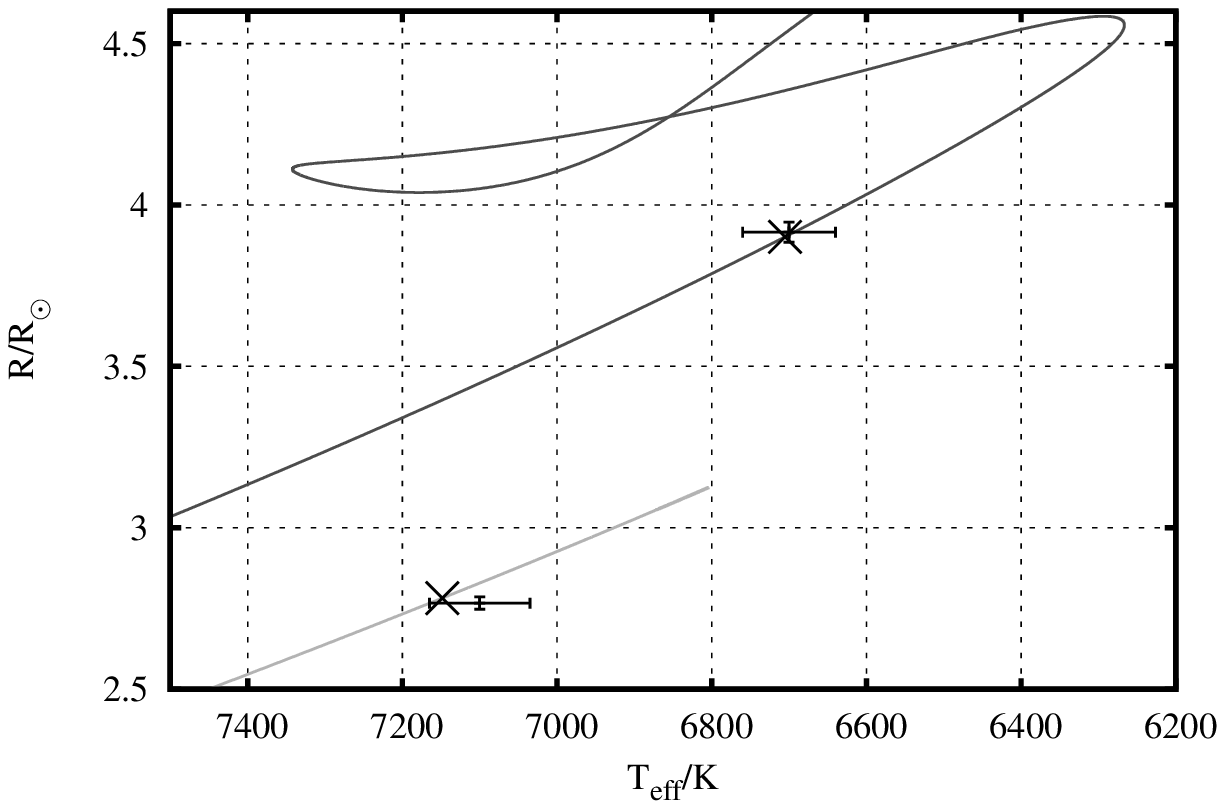}
\includegraphics[width=\columnwidth]{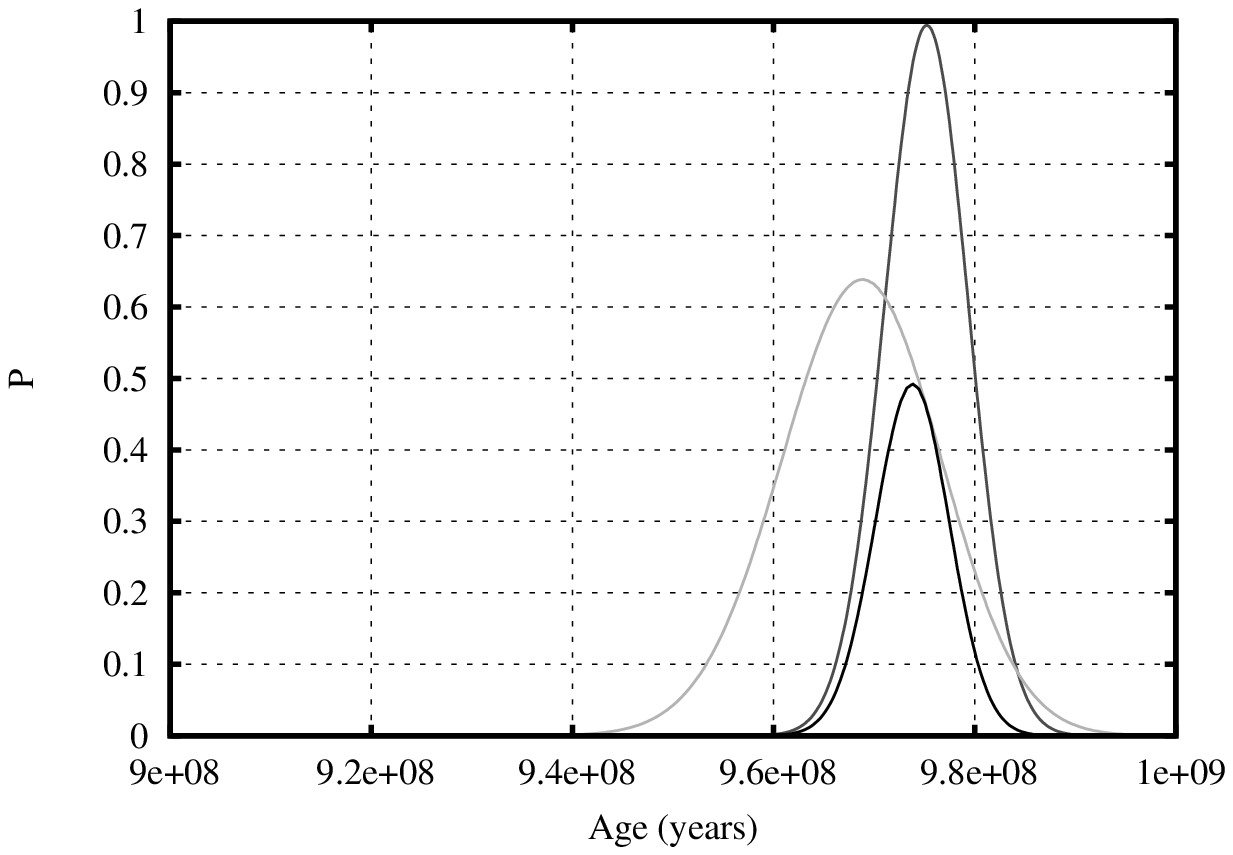}
\includegraphics[width=\columnwidth]{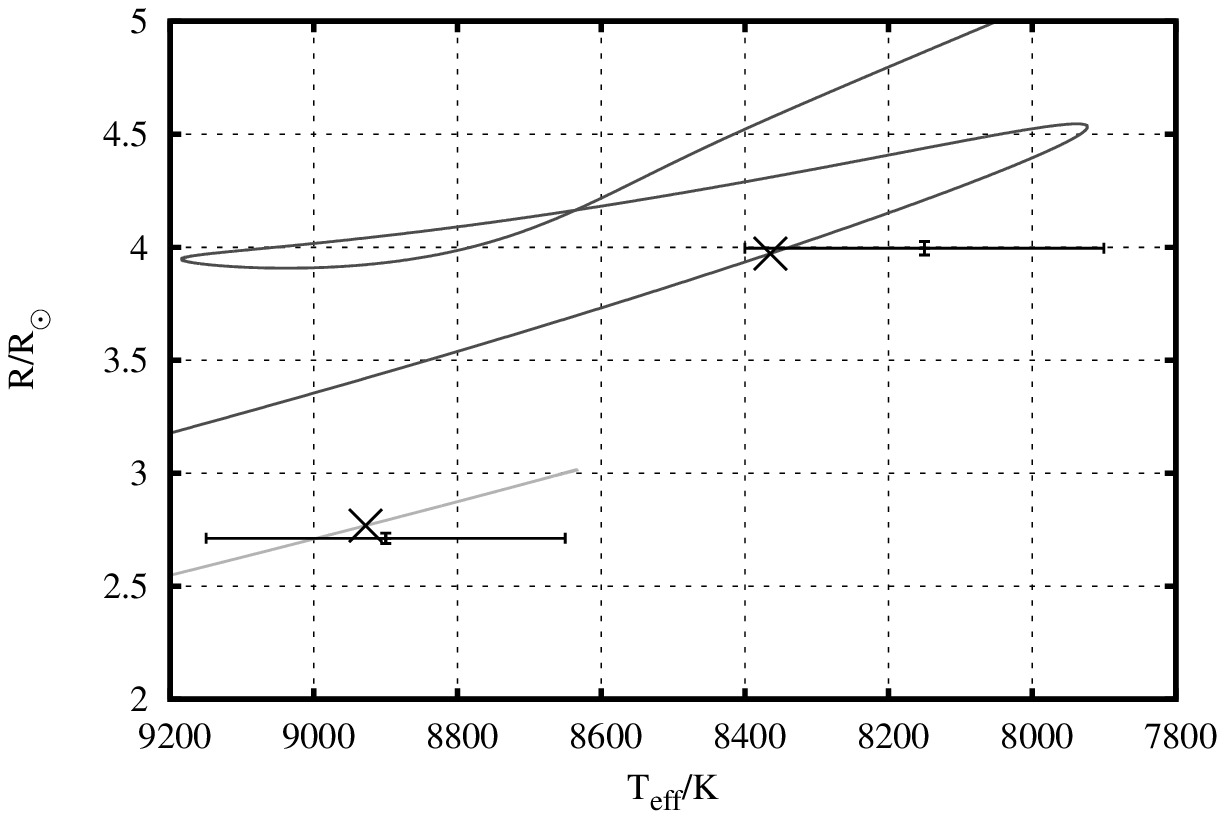}
\includegraphics[width=\columnwidth]{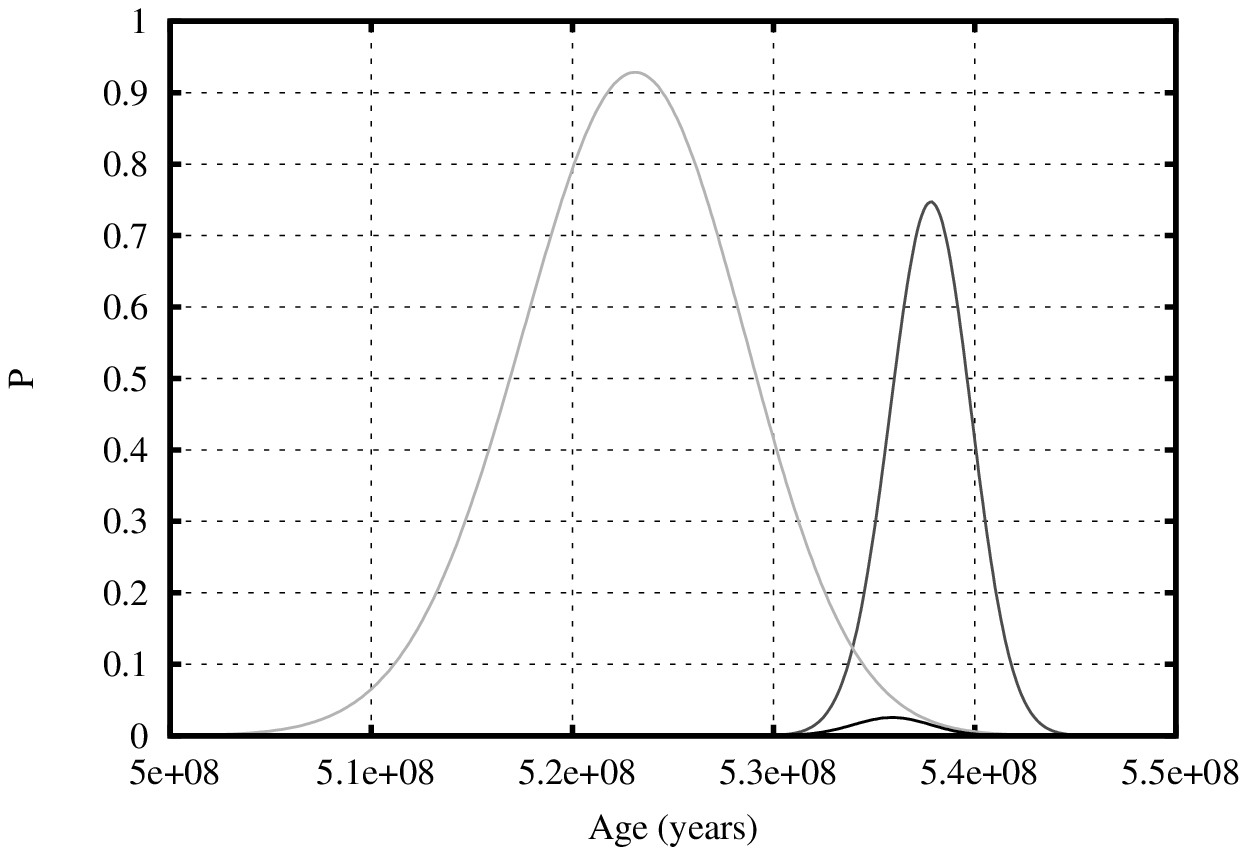}
\end{center}
\caption{{\bf Left column:} Evolutionary tracks in the radius-T$_\mathrm{eff}$ plane for the systems EI Cep (top), AI Hya (middle) and WX Cep (bottom). The primary is denoted by the dark grey line and the secondary by the light grey line. Crosses represent the best fit models for each track and the errorbars denote the observed system. {\bf Right column:} P-values for the systems as a function of age for the systems EI Cep (top), AI Hya (middle) and WX Cep (bottom). The primary is denoted by the dark grey line and the secondary is denoted by the light grey line. The black line is the P-value for the system as a whole.}
\label{fig:RT_P}
\end{figure*}

For this system, we have also computed models for metallicities of Z=0.015 and Z=0.025 in order to get some idea of how sensitive \deltaov\ is to changes in metallicity. The best fit probability falls off rapidly as metallicity of the models is decreased. At Z=0.015, P = $3.292\times10^{-2}$, whereas the decline is much slower at higher metallicites, falling from P = $0.6880$ at Z=0.025 to P = 0.3929 at Z=0.03. The reason for this is shown in Fig.~3. Overshooting has two effects: primarily it extends the length of the main sequence, but it also makes the star slightly more luminous at a given temperature. Lower metallicities favour lower values of \deltaov\ because these models are already hotter and more luminous. The hook at the end of the main sequence occurs at higher temperatures at lower metallicity and for sufficiently small \deltaov, the observed temperatures cannot be reached while the star is on the main sequence.

\begin{figure}
\begin{center}
\includegraphics[angle=270,width=\columnwidth]{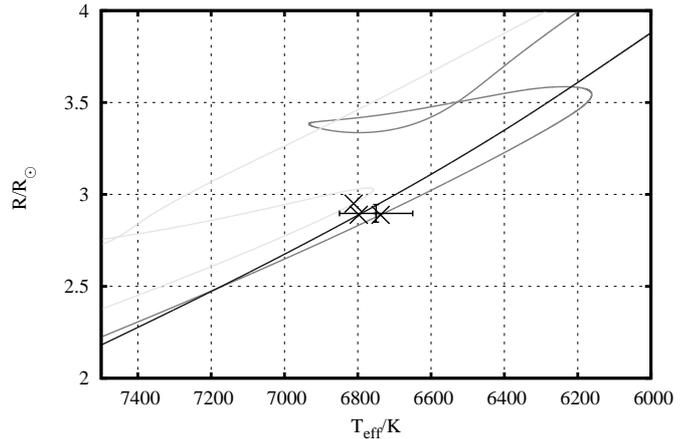}
\caption{Evolutionary tracks in the radius-T$_\mathrm{eff}$ plane for the primary star of EI Cep computed with different metallicities: Z=0.015 (light grey), 0.02 (medium grey) and 0.025 (dark grey). Crosses represent the best fit models for each track and the errorbars denote the observed properties of the system.}
\end{center}
\label{fig:EICepPrimary}
\end{figure}

As an additional test for this system, we also compute models where the initial helium content is varied by $\pm0.05$ (with a corresponding variation in the hydrogen abundance -- i.e. the metal abundance, Z, is held constant). By reducing (increasing) the helium abundance, a model with the same Z and \deltaov\ becomes larger (smaller) at a given temperature but the effect is quite small. This means that models with lower helium abundance tend to require  smaller values of \deltaov. For Z=0.02, a model with Y = 0.275 gives a best fit at \deltaov = 0.12 with P = 0.8595, whereas one with Y = 0.285 gives \deltaov = 0.18 and P = 0.9092. Note that this is a marginally better fit than our standard case.

\subsubsection{Comparison of evolution codes}

In addition to our {\sc stars} models, we have also computed a set of models using the stellar evolution code {\sc mesa} \citep{2011ApJS..192....3P}. Following the prescription of \citet{2000A&A...360..952H}, convective overshooting in this code is implemented by means of a diffusive exponential formalism whereby the extent of mixing is computed via the equation
\be
	D_\mathrm{OV} = D_0 \exp\left(- {2z\over f \mathrm{H_p}}\right),
\ee
where $D_0$ is the MLT diffusion coefficient inside the convective region, $z$ is the distance from the convective boundary, $\mathrm{H_p}$ is the pressure scale height at the convective boundary and $f$ is a dimensionless free parameter. Motivated by the work from \cite{2006ApJS..162..375V},  the overshooting parameter follows a ramp equation:

\begin{equation}
	f = \frac{f_0}{2} \left[  1 - \cos\left(\pi {\frac{M_* - M_{{\rm min}}}{M_{{\rm max}} - M_{{\rm min}}}}\right)   \right]
\end{equation}

\noindent where $f_0$ is a constant, $M_*$ is the current stellar mass of the model, $M_{{\rm min}}$ is the stellar mass below which overshoot mixing does not occur, and $M_{{\rm max}}$ the stellar mass above which $f = f_0$. \cite{2006ApJS..162..375V} gives $M_{{\rm min}} = 1.1$\ms\ and $M_{{\rm max}} = 1.8$\ms. Previously, a parameter of $f_0 \simeq 0.014$ was found to reproduce the width of the main sequence \citep[for more details, see][]{2000A&A...360..952H}.

The {\sc mesa} model grid is run over the same metallicity range as the \stars\ grid (namely Z=0.01, 0.02 and 0.03) and the free parameter $f_0$ is varied from 0 to 0.05 in steps of 0.005.

\begin{figure}
\includegraphics[angle=270, width=\columnwidth]{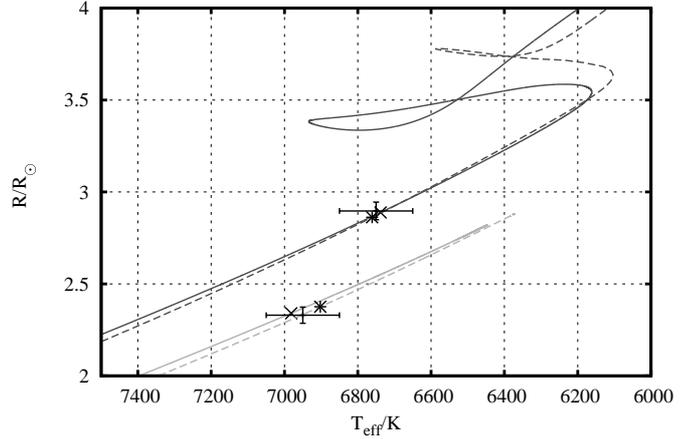}
\caption{Evolutionary tracks in the radius-T$_\mathrm{eff}$ plane for EI Cep. Solid lines denote models computed with \stars, while dashed lines denote models computed with {\sc mesa}. Dark grey denotes the primary and light grey the secondary. The best fit \stars\ model is denoted with crosses, while the best fit {\sc mesa} model is denoted with stars. The black errorbars denote the observed properties of the system.}
\label{fig:STARSvMESA}
\end{figure}

\begin{figure}
\includegraphics[angle=270, width=\columnwidth]{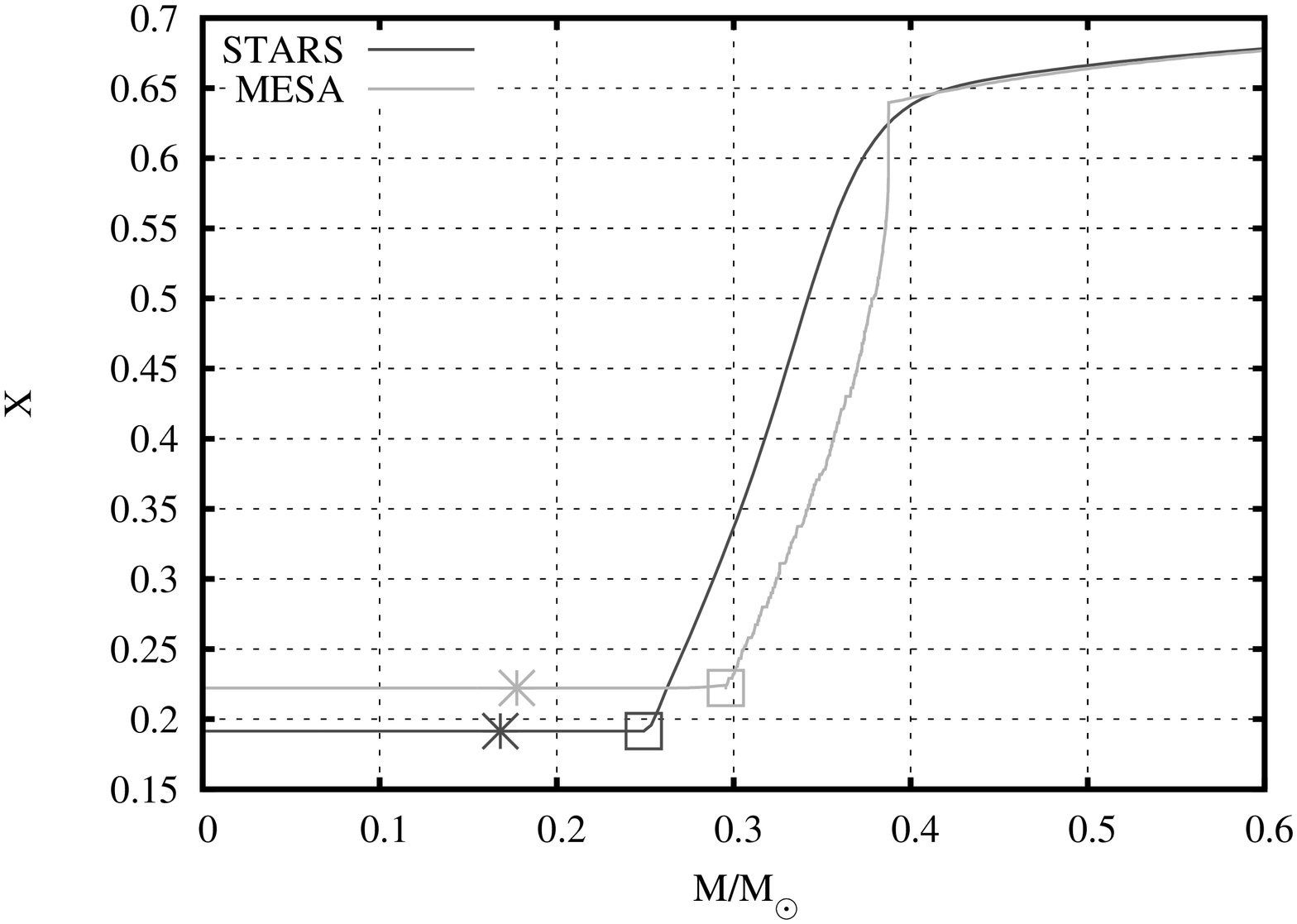}
\includegraphics[angle=270, width=\columnwidth]{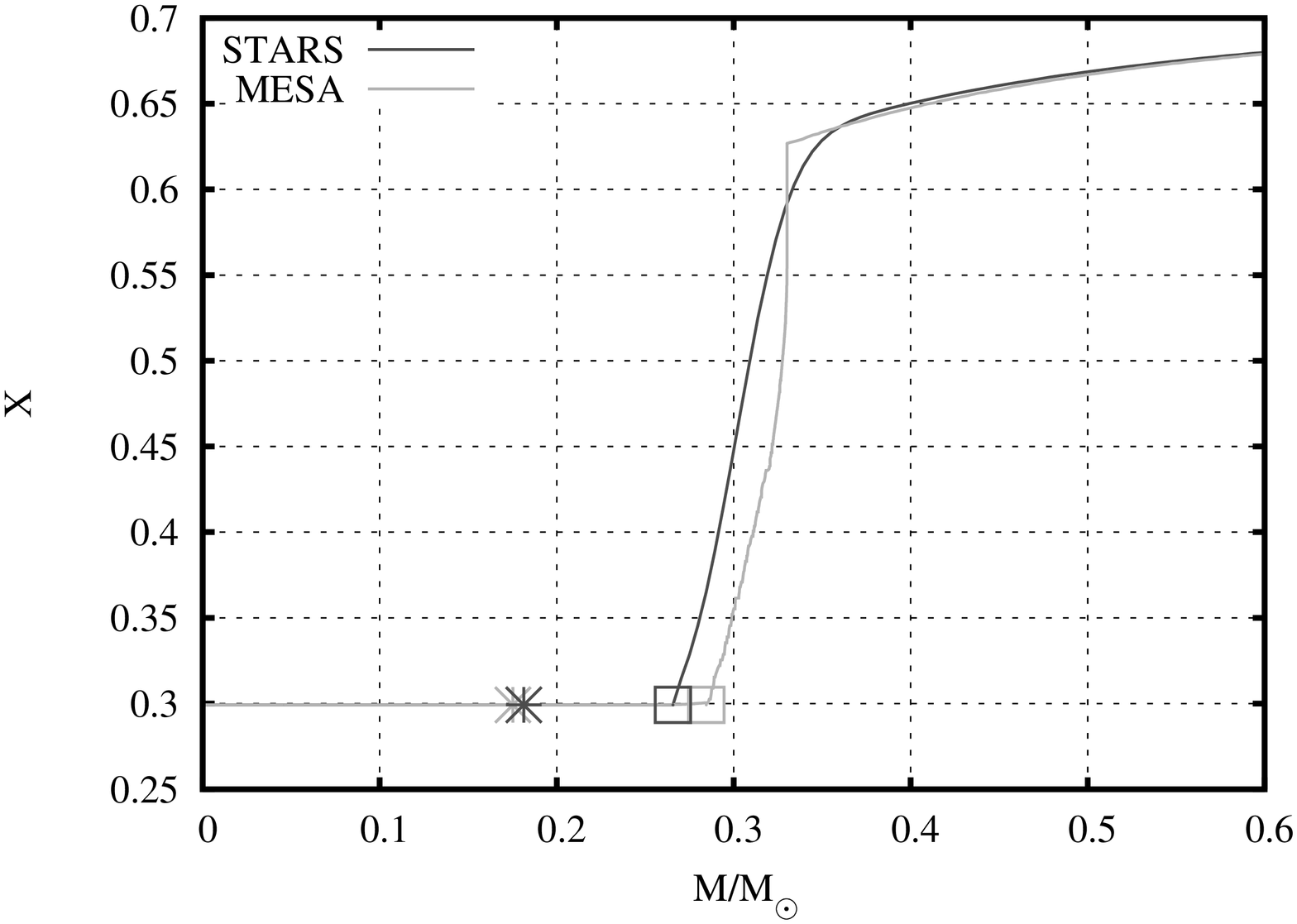}
\caption{Hydrogen profiles as a function of mass for the best fit models for EI Cep. The upper panel shows the primary star and the lower panel the secondary star. Dark grey lines represent the \stars\ model, while the {\sc mesa} model is in light grey. The star represents the edge of the (Schwarzchild) convective core and the square the edge of the overshooting region.}
\label{fig:EICepStructure}
\end{figure}

The best fit model for EI Cep as computed with {\sc mesa} is for $Z=0.02$ and $f_0=0.04$, with P = 0.399. As with the \stars\ models, the quality of fit for the models falls off rapidly as the metallicity is changed such that no good fit is obtained at either $Z=0.01$ or $0.03$. In Fig.~\ref{fig:STARSvMESA}, we show the evolutionary tracks for the best fit models of both {\sc mesa} and \stars. The internal structure of both stars for the best fitting models is displayed in Fig.~\ref{fig:EICepStructure}. The {\sc mesa} model has a larger overshooting region than the \stars\ model which is entirely consistent with the fact that the {\sc mesa} evolutionary track has an end to the main sequence that is redward and brighter than the end of the main sequence in the \stars\ track. If we express the extent of the overshooting region as a fraction of the pressure scale height at the convective boundary, the primary and secondary for the \stars\ model have $f_\mathrm{H_p}=0.251$ and $0.244$, respectively. For the {\sc mesa} models, we obtain $f_\mathrm{H_p}=0.350$ and $0.306$ respectively. The \stars\ hydrogen profile shows much smoother features at the edge of the zone from which the fully mixed region has retreated. This is not due to any physical cause, but is a computational artefact. The \stars\ code employs a non-Lagrangian mesh which is prone to numerical diffusion which acts to smooth out any composition discontinuities \citep[for further details, see][]{2006MNRAS.370.1817S}. Above this region, the two models show very similar structure.

\subsection{V1031 Ori}

For V1031 Ori, our best fit system has a metallicity of Z=0.03. Both \deltaov\ = 0.21 and 0.24 give the same P-value, namely 0.9297 and it is presumed that the true best fit lies between these two values. \citet{2007A&A...475.1019C} gives a best fit for Z=0.015 and 0.15 pressure scale heights of overshooting. \citet{2000MNRAS.318L..55R} find similar parameters, with Z=0.016, Y=0.25 and over 0.2 pressure scale heights of overshooting. \citet{1997MNRAS.289..869P}, on the other hand, prefer a model with overshooting and a metallicity of 0.023, though their model without overshooting and with a metallicity of Z=0.029 is also a good fit. It is difficult to reconcile these different models for the same system.

\subsection{SZ Cen}

For SZ Cen, we obtain our best fit for Z=0.01 and \deltaov=0.12, with P$ = 0.7932$. This places the primary beyond the end of the main sequence. \citet{1997MNRAS.289..869P} give a best fit metallicity between 0.018 and 0.024, while \citet{2007A&A...475.1019C} obtains a best fit at Z=0.02 and with 0.1 pressure scale heights of overshooting. However, it should be noted that \citet{2010A&ARv..18...67T} revised the parameters for this system, compared to the values used in both of the above works. They adopted a temperature based on the beta index, rather than on an assumed spectral type coupled with a colour-T$_\mathrm{eff}$ calibration (G. Torres, private communication). The result of this is that the temperatures given by \citet{2010A&ARv..18...67T} are somewhat hotter than those used in earlier works. If we adopt the parameters for SZ Cen used by \citet{2007A&A...475.1019C}, we obtain a best fit at Z=0.03 for \deltaov\ = 0.03 with P = 0.0121. This does not represent a good fit. On the other hand, \citet{2000MNRAS.318L..55R} obtain a best fit to this system with Z=0.007, Y=0.2 and 0.1-0.2 pressure scale heights of overshooting. Aside from the low helium value, our result is consistent with these parameters.

\begin{figure}
\begin{center}
\includegraphics[width=\columnwidth]{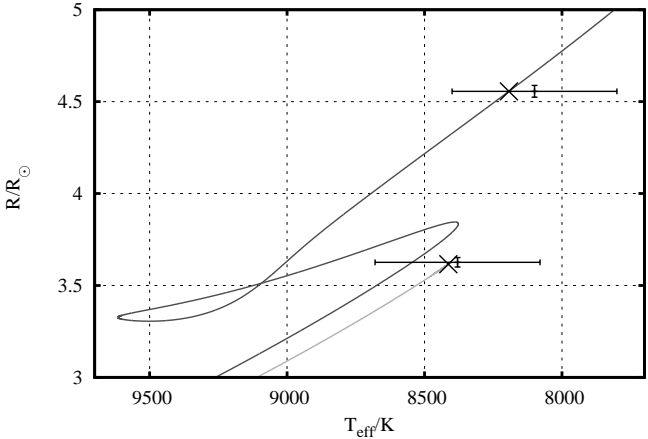}
\caption{Evolutionary tracks in the radius-T$_\mathrm{eff}$ plane for the components of SZ Cen. The primary is denoted by the dark grey line and the secondary by the light grey line. Crosses represent the best fit models for each track and the errorbars denote the observed system.}
\end{center}
\label{fig:SZCenRT}
\end{figure}

\subsection{AY Cam, CV Vel and V539 Ara}

These three systems all give excellent fits. For AY Cam, we obtain a best fit for Z=0.02 and \deltaov\ = 0.09, with a P-value of 0.9267. We have been unable to find any other attempts to fit this system in the literature. CV Vel and V539 Ara are considerably more massive than the other binaries in our sample and both have primary masses around 6\ms. For both systems, we find a best fit at Z=0.02, though the \deltaov\ values are different for both. For CV Vel, we prefer \deltaov\ = 0.09, while for V539 Ara \deltaov\ = 0.15 gives the best fit. \citet{1997MNRAS.289..869P} also attempted to fit both these systems. For CV Vel, they give a slight preference to models without overshooting and suggest the best fit metallicity is Z=0.016. For V539 Ara, their overshooting models are preferred and Z=0.016 is again the best fit metallicity.

\subsection{AI Hya, TZ For and V2291 Oph}

Though not included in the Torres et al. list, we have also attempted to model the system V2291 Oph. This $\zeta$ Auriga system was used by \citet{1997MNRAS.285..696S} for their overshooting calibration and it was a key system in their determination of the overshooting parameter. The parameters of this system are listed in Table~\ref{tab:V2291Oph}.

\begin{table}
\begin{center}
\begin{tabular}{lcc}
& Primary & Secondary \\
\hline
Mass (\ms) & 3.86$\pm$0.15 & 2.95$\pm$0.09\\
Radius (R$_\odot$) & 32.9$\pm$1.5 & 3.0$\pm$0.2 \\
T$_\mathrm{eff}$ (K) & 4\,850$\pm$100 & 11\,000$\pm$500 \\
\hline
\end{tabular}
\end{center}
\caption{Parameters of the system V2291 Oph. These data are taken from \citet{1997MNRAS.285..696S}.}
\label{tab:V2291Oph}
\end{table} 

We are able to obtain reasonable, if not excellent, fits to the systems AI Hya, TZ For and V2291 Oph. These latter two systems are significantly evolved. \citet{1997MNRAS.285..696S} obtained \deltaov\ = 0.12, for a model with X=0.70, Y=0.28 and Z=0.02. \citet{2009A&A...507..377C} chose a higher metallicity based on chemical analysis by \citet{1996MNRAS.280..977M}, preferring Z=0.03, and finds a best fit when 0.2 pressure scale heights of overshooting are used. Our best fit is for $Z=0.03$ and \deltaov$ = 0.24$, with P = 0.4777. This places the primary on the first ascent of the giant branch. However,  we also obtain a fit of P = 0.2240 for $Z=0.02$ and \deltaov$=0.15$ which would place the primary in the core helium burning phase, in agreement with the Pols et al. solution.

TZ For is another system in which the primary is highly evolved. For this system, we obtain a best fit for $Z=0.03$ and \deltaov\ =0.24, with P = 0.7036. This places the primary on the first ascent of the red giant branch. The P-values for neighbouring \deltaov\ and Z-values in our grid are all less than 0.1, so this solution is strongly favoured. Previous attempts to obtain a solution for this system have also had issues. \citet{2007A&A...475.1019C} suggests using 0.6 pressure scale heights of overshooting, but notes that this seems to be too large for stars of masses comparable to TZ For. \citet{1997MNRAS.289..869P} find slight evidence in favour of models with overshooting, but their solution is also less than ideal (see the right-hand panel of their figure 11). 

Unlike the previous two systems, AI Hya has both its components on the main sequence. For this system, we find a best fit for Z=0.04 and \deltaov\ =0.15. Fits from neighbouring metallicities are notably worse, as are neighbouring \deltaov\ values. Our fit is not perfect mostly because we do not find a good fit to the secondary, as shown in the middle-right panel of Figure~\ref{fig:RT_P}. Our solution is in good agreement with \citet{1997MNRAS.289..869P}, who found clear evidence for a superior solution when overshooting was included, but noted that the best fit metallicity lay outside the metallicity range (their maximum Z was 0.033) of their isochrones. \citet{2007A&A...475.1019C} attempted to fit this system with models at Z=0.02 and was unable to obtain a satisfactory solution.

\subsection{V364 Lac, WX Cep, AQ Ser}

For three of our systems, we are unable to obtain reasonable fits at all. For V364 Lac, WX Cep and AQ Ser, the best-fit P-value is below 0.1.

Of the three systems, WX Cep is the least worst fitting system. We find a best fit value of P = 0.025 at Z=0.02 and \deltaov\ = 0.12. The main difficulty with this system is that the best fits to the components do not occur at the same age. The fit to the primary alone is at a maximum of about 0.75 at an age of $5.37\times10^8$\,yr, but the fit to the secondary reaches a peak value of 0.92 at just $5.22\times10^8$\,yr. There is little overlap between the two fit distributions, resulting in a poor fit overall, as can be seen in Fig.~\ref{fig:RT_P}. Our best-fit metallicity (and hence also the age of the system) agrees with that of \citet{2007A&A...475.1019C}, who finds a best-fit solution for 0-0.2 pressure scale heights of overshooting. \citet{1997MNRAS.289..869P} also favour a model with overshooting and a metallicity of Z=0.02.

V364 Lac is a peculiar case. We are able to obtain almost perfect fits to the individual components of this system (with best-fit values of almost unity) at Z=0.02 and with \deltaov\ = 0.09. However, these solutions have very different ages. The primary is best fit at an age of $6.7\times10^8$\,yr, while the secondary is best fit at an age of $5.9\times10^8$\,yr. We are not aware of any other attempts to fit this system in the literature. 

\begin{figure}
\includegraphics[width=\columnwidth]{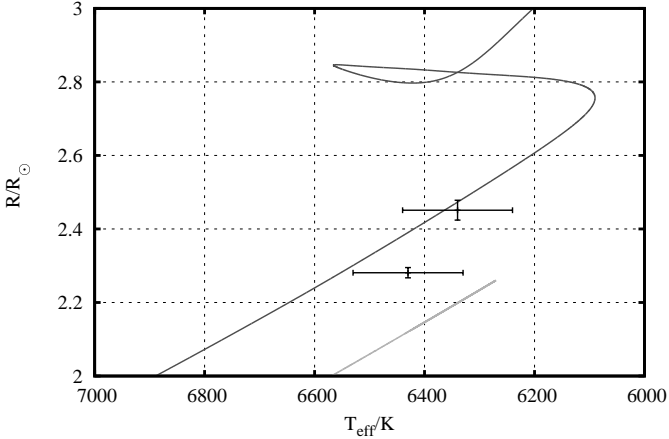}
\caption{Evolutionary tracks in the radius-T$_\mathrm{eff}$ plane for the components of AQ Ser, assuming a model with Z = 0.01 and \deltaov\ = 0.18. The primary is denoted by the dark grey line and the secondary by the light grey line. The errorbars denote the observed system.}
\label{fig:AQSer_RT}
\end{figure}

AQ Ser has already been noted as a problematic system by \citet{2014AJ....147...36T}. These authors note that the more massive star appears systematically younger than its less massive counterpart. The P-value for our best fit to AQ Ser is so vanishingly small that it may well be considered zero. We can obtain a reasonable fit to the primary alone with Z=0.01 and \deltaov\ = 0.18 (see Fig.~\ref{fig:AQSer_RT}), but it is clear from the evolutionary track that our secondary is not massive enough. It is possible that a better fit to the system would be obtained if the masses of the two components were closer together, given how similar the radii and effective temperatures of the two stars are. As a test of this, we ran a single calculation with Z=0.01 and \deltaov\ = 0.18, in which we set the primary mass to be 1.395\ms\ and the secondary mass to be 1.370\ms\ which are the minimum and maximum masses (respectively) allowed by the errors. In this case, we obtain a fit of P=0.101, which is a dramatic improvement over our original calculations.

\section{Discussion and conclusions}

\begin{figure}
\includegraphics[width=\columnwidth]{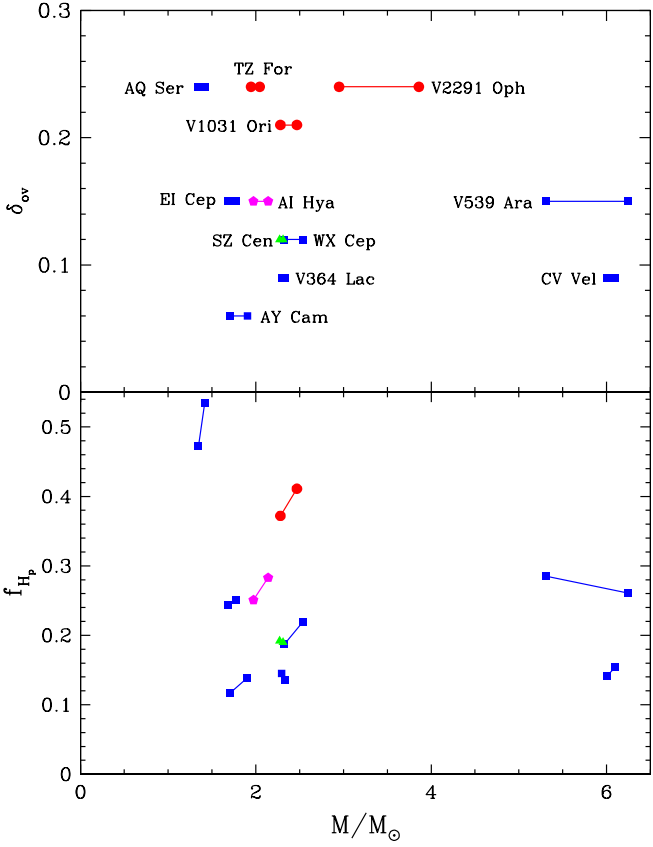}
\caption{Extent of overshooting as a function of mass. The upper panel displays the overshooting in terms of \deltaov, while the lower panel shows the same quantity in terms of a fraction of the pressure scale height. Symbols denote the best fit metallicity for the systems: Z=0.01 (triangles), Z=0.02 (squares), Z=0.03 (circles) and Z=0.04 (pentagons).}
\label{fig:MassOS}
\end{figure}

We summarise the best fits to each of our systems in Table~\ref{tab:RTbestfit}. In Fig.~\ref{fig:MassOS} we have plotted the extent of overshooting as a function of stellar mass, showing the overshooting both in terms of \deltaov\ and as a fraction of the pressure scale height. Note that we have not plotted TZ For and V2291 Oph on the pressure scale height plot, as these systems are evolved beyond the main sequence and we cannot define this quantity if there is no longer a convective core. For all of our systems, we require some degree of overshooting and in agreement with previous works the amount required is only moderate. The extent of overshooting ranges from \deltaov\ =\ 0.06 to 0.24, though the highest values are found only in our most evolved and most problematic systems (AQ Ser, TZ For and V2291 Oph). Most systems have \deltaov\ between 0.09 and 0.15. This is consistent with the 0.1-0.3 pressure scale heights of overshooting found by other authors. We see no reason to suggest that \deltaov\ is a function of mass \citep[unlike][]{2000MNRAS.318L..55R}, though we note that the majority of our systems are clustered around 2\ms, with just two systems at around 6\ms.  There is also no evidence for a metallicity dependence to the extent of overshooting, though this may simply be because we lack enough systems across a range of metallicities. While our three Z=0.03 systems do have higher \deltaov\ values, we stress that two of these are very poor fits hence we cannot provide a definite conclusion regarding a possible metallicity dependence.

The extent of overshooting we find is consistent with values obtained from asteroseismic determinations. \citet{aerts2015} finds values between 0-0.5 pressure scale heights, also with considerable star-to-star variations. She also reports that there is no obvious relation between extent of overshooting and stellar mass. \citet{2012A&A...539A..90N} find 0.3-0.35 pressure scale heights of overshooting in their two late Be stars, HD 181231 and HD 175869. Again our results are consistent with these values.

To complement our {\sc stars} models, we have also calculated best fit models for each of our systems using {\sc mesa}. The details of these fits are given in the appendix. We find general agreement between the two codes. Both predict the same metallicity for all but one of the systems (the exception is AI Hya). There is also agreement in the trend of the overshooting parameter required: for systems where {\sc stars} requires a high \deltaov, we also find that {\sc mesa} requires a high $f_0$. The extent of overshooting in the {\sc mesa} models is about 0.15-0.4 pressure scale heights. Computing a P-weighted average of the overshooting parameter $f_0$ results in a value of 0.031.

\begin{table}
\begin{tabular}{lccc}
System & Best fit Z & Best fit \deltaov & P \\
\hline
V364 Lac & 0.02 & 0.09 &  $6.914\times10^{-4}$ \\
AI Hya & 0.04 & 0.15 &  0.4922 \\
EI Cep & 0.02 & 0.15 &  0.9066 \\
TZ For & 0.03 & 0.24 &  0.7036 \\
WX Cep & 0.02 & 0.12 & $2.549\times10^{-2}$ \\
V1031 Ori & 0.03 & 0.21(4) &  0.9297 \\
SZ Cen & 0.01 &  0.12 & 0.8767  \\
AY Cam & 0.02 & 0.09 & 0.9267  \\
AQ Ser & 0.02 & 0.24 & $8.501\times10^{-55}$\\
V2291 Oph & 0.03 & 0.24 & 0.4777 \\
CV Vel & 0.02 & 0.09 & 0.8632 \\
V539 Ara & 0.02 & 0.15 & 0.9267 \\
\hline
\end{tabular}
\caption{Best fit properties for each of our systems using only T and R.}
\label{tab:RTbestfit}
\end{table}

Ideally, one would like to be able to select a single value for the overshooting parameter that would work reasonably well for all systems. This is particularly relevant for future work in our project, where we aim to create grids of low-mass stellar models for use in {\sc bonnsai}. If we neglect the systems with P-values less than 0.1 and take an average of the overshooting parameter for the remaining systems, in which we weight each system by its relative P-value, we obtain \deltaov\ = 0.156. Adopting this value, we would still obtain good fits for the systems CV Vel and V1031 Ori (which give P=0.7551 and 0.7243 respectively for \deltaov\ = 0.15). AY Cam and V2291 Oph would be moderate fits (P = 0.2733 and 0.2240 respectively), though the best fit metallicity for the latter system would drop to Z = 0.02. For the systems SZ Cen and TZ For we are unable to obtain any fit at \deltaov\ = 0.15. Of the nine systems we are able to fit when our choice of Z and \deltaov\ is free, we would only fit seven of these systems if we adopted our average value of \deltaov = 0.156.

It would be useful to be able to extend this work to a wider sample of binaries, particularly as many of our systems cluster around 2\ms. Additional data for stars of between 3 and 5 \ms\ (and above 6\ms) would be particularly desirable. However, as can be seen from Fig.~\ref{fig:selected_hr}, few systems in the \citet{2010A&ARv..18...67T} data set lie in this mass range even before we take into account our requirements that the systems should be fairly wide and contain components that are evolved away from the ZAMS. Nature, it seems, has not been kind to us in this regard. One can hope that future survey missions, such as Gaia\footnote{\texttt{sci.esa.int/gaia/}}, may help to fill this deficiency.

\begin{acknowledgements}

The authors thank G. Torres for useful discussion regarding the parameters of SZ Cen and the referee for her/his useful remarks. RJS is the recipient of a Sofja Kovalevskaja Award from the Alexander von Humboldt Foundation. LF and JCP acknowledge funding from the Alexander von Humboldt Foundation. FNRS is funded via the Bonn Cologne Graduate School. JCP also thanks Norbert Langer for his support.

\end{acknowledgements}

\bibliographystyle{aa}
\bibliography{/Users/richardstancliffe/Work/NewBib}

\begin{thebibliography}{28}
\expandafter\ifx\csname natexlab\endcsname\relax\def\natexlab#1{#1}\fi

\bibitem[{{Aerts}(2015)}]{aerts2015}
{Aerts}, C. 2015, in Proceedings of IAU symposium 307, ed. J.~G. . P.~S.
  G.~Meynet, C.~Georgy (Cambridge University Press), in press (arXiv:1407.6479)

\bibitem[{{Andersen}(1991)}]{1991A&ARv...3...91A}
{Andersen}, J. 1991, \aapr, 3, 91

\bibitem[{{Brott} {et~al.}(2011){Brott}, {de Mink}, {Cantiello}, {Langer}, {de
  Koter}, {Evans}, {Hunter}, {Trundle}, \& {Vink}}]{2011A&A...530A.115B}
{Brott}, I., {de Mink}, S.~E., {Cantiello}, M., {et~al.} 2011, \aap, 530, A115

\bibitem[{{Casagrande} {et~al.}(2011){Casagrande}, {Sch{\"o}nrich}, {Asplund},
  {Cassisi}, {Ram{\'{\i}}rez}, {Mel{\'e}ndez}, {Bensby}, \&
  {Feltzing}}]{2011A&A...530A.138C}
{Casagrande}, L., {Sch{\"o}nrich}, R., {Asplund}, M., {et~al.} 2011, \aap, 530,
  A138

\bibitem[{{Claret}(2007)}]{2007A&A...475.1019C}
{Claret}, A. 2007, \aap, 475, 1019

\bibitem[{{Claret}(2009)}]{2009A&A...507..377C}
{Claret}, A. 2009, \aap, 507, 377

\bibitem[{{Eggleton}(1971)}]{1971MNRAS.151..351E}
{Eggleton}, P.~P. 1971, \mnras, 151, 351

\bibitem[{{Eggleton}(1972)}]{1972MNRAS.156..361E}
{Eggleton}, P.~P. 1972, \mnras, 156, 361

\bibitem[{{Guenther} {et~al.}(2014){Guenther}, {Demarque}, \&
  {Gruberbauer}}]{2014ApJ...787..164G}
{Guenther}, D.~B., {Demarque}, P., \& {Gruberbauer}, M. 2014, \apj, 787, 164

\bibitem[{{Herwig}(2000)}]{2000A&A...360..952H}
{Herwig}, F. 2000, \aap, 360, 952

\bibitem[{{K{\"o}hler} {et~al.}(2015){K{\"o}hler}, {Langer}, {de Koter}, {de
  Mink}, {Crowther}, {Evans}, {Gr{\"a}fener}, {Sana}, {Sanyal}, {Schneider}, \&
  {Vink}}]{2015A&A...573A..71K}
{K{\"o}hler}, K., {Langer}, N., {de Koter}, A., {et~al.} 2015, \aap, 573, A71

\bibitem[{{Maeder} \& {Meynet}(1991)}]{1991A&AS...89..451M}
{Maeder}, A. \& {Meynet}, G. 1991, \aaps, 89, 451

\bibitem[{{Marshall}(1996)}]{1996MNRAS.280..977M}
{Marshall}, K.~P. 1996, \mnras, 280, 977

\bibitem[{{Meng} \& {Zhang}(2014)}]{2014ApJ...787..127M}
{Meng}, Y. \& {Zhang}, Q.~S. 2014, \apj, 787, 127

\bibitem[{{Montalb{\'a}n} {et~al.}(2013){Montalb{\'a}n}, {Miglio}, {Noels},
  {Dupret}, {Scuflaire}, \& {Ventura}}]{2013ApJ...766..118M}
{Montalb{\'a}n}, J., {Miglio}, A., {Noels}, A., {et~al.} 2013, \apj, 766, 118

\bibitem[{{Neiner} {et~al.}(2012){Neiner}, {Mathis}, {Saio}, {Lovekin},
  {Eggenberger}, \& {Lee}}]{2012A&A...539A..90N}
{Neiner}, C., {Mathis}, S., {Saio}, H., {et~al.} 2012, \aap, 539, A90

\bibitem[{{Paxton} {et~al.}(2011){Paxton}, {Bildsten}, {Dotter}, {Herwig},
  {Lesaffre}, \& {Timmes}}]{2011ApJS..192....3P}
{Paxton}, B., {Bildsten}, L., {Dotter}, A., {et~al.} 2011, \apjs, 192, 3

\bibitem[{{Pols} {et~al.}(1995){Pols}, {Tout}, {Eggleton}, \&
  {Han}}]{1995MNRAS.274..964P}
{Pols}, O.~R., {Tout}, C.~A., {Eggleton}, P.~P., \& {Han}, Z. 1995, \mnras,
  274, 964

\bibitem[{{Pols} {et~al.}(1997){Pols}, {Tout}, {Schr\"oder}, {Eggleton}, \&
  {Manners}}]{1997MNRAS.289..869P}
{Pols}, O.~R., {Tout}, C.~A., {Schr\"oder}, K.-P., {Eggleton}, P.~P., \&
  {Manners}, J. 1997, \mnras, 289, 869

\bibitem[{{Ribas} {et~al.}(2000){Ribas}, {Jordi}, \&
  {Gim{\'e}nez}}]{2000MNRAS.318L..55R}
{Ribas}, I., {Jordi}, C., \& {Gim{\'e}nez}, {\'A}. 2000, \mnras, 318, L55

\bibitem[{{Schneider} {et~al.}(2014){Schneider}, {Langer}, {de Koter}, {Brott},
  {Izzard}, \& {Lau}}]{2014arXiv1408.3409S}
{Schneider}, F.~R.~N., {Langer}, N., {de Koter}, A., {et~al.} 2014, \aap, 570,
  A66

\bibitem[{{Schr\"oder} {et~al.}(1997){Schr\"oder}, {Pols}, \&
  {Eggleton}}]{1997MNRAS.285..696S}
{Schr\"oder}, K.-P., {Pols}, O.~R., \& {Eggleton}, P.~P. 1997, \mnras, 285, 696

\bibitem[{{Stancliffe}(2006)}]{2006MNRAS.370.1817S}
{Stancliffe}, R.~J. 2006, \mnras, 370, 1817

\bibitem[{{Stancliffe} \& {Eldridge}(2009)}]{2009MNRAS.396.1699S}
{Stancliffe}, R.~J. \& {Eldridge}, J.~J. 2009, \mnras, 396, 1699

\bibitem[{{Tkachenko} {et~al.}(2014){Tkachenko}, {Aerts}, {Pavlovski},
  {Degroote}, {P{\'a}pics}, {Moravveji}, {Lehmann}, {Kolbas}, \&
  {Cl{\'e}mer}}]{2014MNRAS.442..616T}
{Tkachenko}, A., {Aerts}, C., {Pavlovski}, K., {et~al.} 2014, \mnras, 442, 616

\bibitem[{{Torres} {et~al.}(2010){Torres}, {Andersen}, \&
  {Gim{\'e}nez}}]{2010A&ARv..18...67T}
{Torres}, G., {Andersen}, J., \& {Gim{\'e}nez}, A. 2010, \aapr, 18, 67

\bibitem[{{Torres} {et~al.}(2014){Torres}, {Vaz}, {Sandberg Lacy}, \&
  {Claret}}]{2014AJ....147...36T}
{Torres}, G., {Vaz}, L.~P.~R., {Sandberg Lacy}, C.~H., \& {Claret}, A. 2014,
  \aj, 147, 36

\bibitem[{{VandenBerg} {et~al.}(2006){VandenBerg}, {Bergbusch}, \&
  {Dowler}}]{2006ApJS..162..375V}
{VandenBerg}, D.~A., {Bergbusch}, P.~A., \& {Dowler}, P.~D. 2006, \apjs, 162,
  375

\end{thebibliography}

\appendix

\section{{\sc mesa} model fits}

\begin{figure}
\includegraphics[width=\columnwidth]{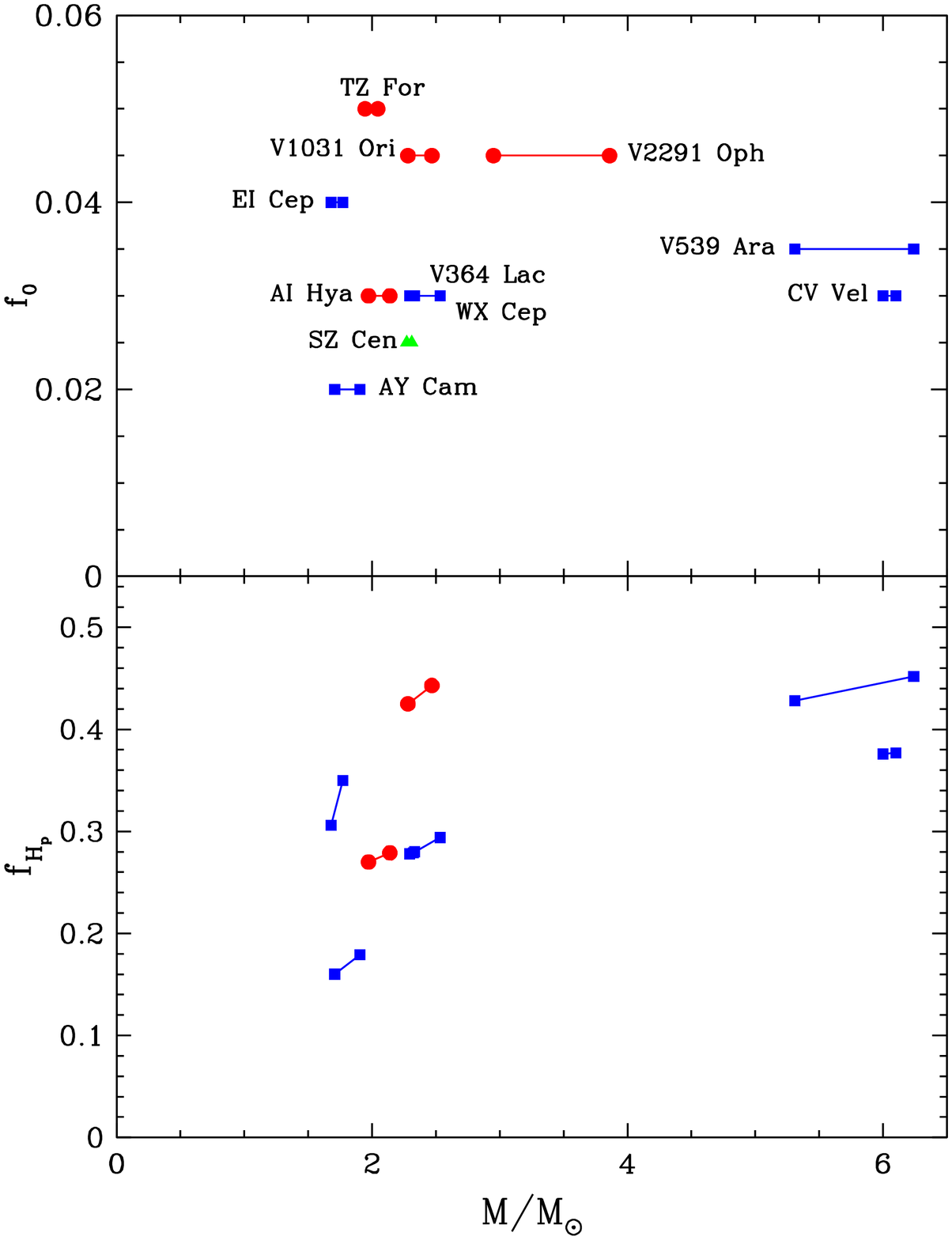}
\caption{Extent of overshooting as a function of mass for the {\sc mesa} models. The upper panel displays the overshooting in terms of \deltaov, while the lower panel shows the same quantity in terms of a fraction of the pressure scale height. Symbols denote the best fit metallicity for the systems: Z=0.01 (triangles), Z=0.02 (squares), Z=0.03 (circles).}
\label{fig:MassOS_mesa}
\end{figure}

In addition to EI Cep, we have also computed models for all our systems using {\sc mesa}. The best fit parameters are given in Table~\ref{tab:RTbestfit_mesa}. Fig.~\ref{fig:MassOS_mesa} shows this data, together with the extent of the overshooting region as a function of the pressure scale height at the top of the convective region. Calculating an average $f_0$ weighted by the P-values, we obtain $f_0 = 0.031$.

\begin{table}
\begin{tabular}{lccc}
System & Best fit Z & Best fit $f_0$ & P \\
\hline
V364 Lac & 0.02 & 0.030 & $2.244\times10^{-3}$ \\
AI Hya & 0.03 & 0.030 & $1.604\times10^{-3}$    \\
EI Cep & 0.02 & 0.040 & 0.3987 \\
TZ For &  0.03 & 0.050 & 0.01072 \\
WX Cep & 0.02 & 0.030 & 0.05469 \\
V1031 Ori &  0.03 & 0.045 & 0.9474 \\
SZ Cen & 0.01 & 0.025 & 0.01779  \\
AY Cam &  0.02 & 0.020 & 0.5296  \\
AQ Ser &  - & - \\
V2291 Oph & 0.03 & 0.045  & 0.09719 \\
CV Vel & 0.02 & 0.030 & 0.9427  \\
V539 Ara & 0.02 & 0.035 & 0.9808  \\
\hline
\end{tabular}
\caption{Best fit properties for each of our systems using only T and R as determined using the {\sc mesa} code. No reasonable fit could be found to AQ Ser (P $< 10^{-96}$).}
\label{tab:RTbestfit_mesa}
\end{table}

\end{document}